\documentclass[twocolumn]{article}

\usepackage[utf8]{inputenc}

\usepackage{microtype}

\usepackage{hyperref}
\usepackage{graphicx}
\usepackage{amsmath,amscd,amsfonts}
\usepackage{siunitx}
\usepackage{color}

\usepackage{tabularx}
\usepackage{xcolor}
\usepackage{authblk}

\begin{document}

\title{Adaptive optics benefit for quantum key distribution uplink from ground to a satellite}
\author[1,2,3,4]{Christopher J. Pugh}
\author[5,6]{Jean-Francois Lavigne}
\author[1,2,7]{Jean-Philippe Bourgoin}
\author[1,2]{Brendon L. Higgins} 
\author[1,2,*]{Thomas Jennewein}

\affil[1]{Institute for Quantum Computing, University of Waterloo, ON, N2L 3G1, Canada}
\affil[2]{Department of Physics and Astronomy, University of Waterloo, ON, N2L 3G1, Canada}
\affil[3]{Department of Physics and Astronomy, Brandon University, Brandon, MB, R7A 6A9, Canada}
\affil[4]{Faculty of Physics, Astronomy and Informatics, Nicolaus Copernicus University, Torun, Poland}
\affil[5]{Institut national d'optique, Quebec, QC, G1P 4S4, Canada}
\affil[6]{Current affiliation: ABB, Quebec, QC, G1P 0B2, Canada}
\affil[7]{Current Affiliation: Aegis Quantum, Waterloo, ON, Canada}
\affil[*]{email: thomas.jennewein@uwaterloo.ca}
\renewcommand\Affilfont{\itshape\small}
\date{}

\maketitle
\begin{abstract}
For quantum communications, the use of Earth-orbiting satellites to extend distances has gained significant attention in recent years, exemplified in particular by the launch of the Micius satellite in 2016. 
The performance of applied protocols such as quantum key distribution (QKD) depends significantly upon the transmission efficiency through the turbulent atmosphere, which is especially challenging for ground-to-satellite uplink scenarios. Adaptive optics (AO) techniques have been used in astronomical, communication, and other applications to reduce the detrimental effects of turbulence for many years, but their applicability to quantum protocols, and their requirements specifically in the uplink scenario, are not well established. Here, we model the effect of the atmosphere on link efficiency between an Earth station and a satellite using an optical uplink, and how AO can help recover from loss due to turbulence. Examining both low-Earth-orbit and geostationary uplink scenarios, we find that a modest link transmissivity improvement of about \SI{3}{\dB} can be obtained in the case of a co-aligned downward beacon, while the link can be dramatically improved, up to \SI{7}{\dB}, using an offset beacon, such as a laser guide star. AO coupled with a laser guide star would thus deliver a significant increase in the secret key generation rate of the QKD ground-to-space uplink system, especially as reductions of channel loss have favourably nonlinear key-rate response within this high-loss regime.
\end{abstract}

\section{Introduction}

For two parties to securely communicate over public channels, they must utilize encryption, consuming shared secret keys in the process.
Whereas classical key generation schemes rely on assumptions of computational complexity, quantum key distribution (QKD) relies on foundational principles of quantum mechanics~\cite{GRT02,SBC09}. Practical QKD implementations depend on the transmission of quantum optical signals either through optical fiber or free space~\cite{BBBSS92, FI94}. In both cases, however, the fragility of a single photon channel fundamentally limits the distances that can be achieved: losses in fiber restrict the maximum distance to a few hundred kilometers~\cite{KLH15, YCY16}, while terrestrial free-space implementations are limited by line-of-sight and thus the curvature of Earth (distances up to \SI{144}{\km} have been demonstrated~\cite{FWS06, UTSWSLBJPTOFMRSBWZ07}).

QKD can be scaled up to global distances by using orbiting satellites acting as intermediate nodes between two ground stations~\cite{NHM02,AJP03,RGK04,AFJ08}. With a ``trusted'' node~\cite{JBH14}, the satellite combines keys generated by separate QKD links to each ground station---effectively encrypting one key with the other---and transmits the result to ground. One ground station then uses its own key to extract the other key (from the combined result), which can be used to encrypt messages. The satellite is trusted in the sense that it has access to each ground station's key in this process. Alternatively, the satellite could be an ``untrusted'' node~\cite{UJZ09} by providing entangled photons to both ground stations simultaneously, who can together verify their integrity. Both of these approaches were demonstrated recently with the Micius satellite~\cite{LCL17,LCH18}.

With the trusted node approach, either an optical uplink (ground to satellite) or downlink (satellite to ground) is possible, both of which are under active investigation~\cite{UJKPCMAVSABBBCCGLHLLMPRRRSSTTTOPVWWWZZ08, X11, TTTKA11, BMHHEHKHDGLJ13,X16}. With all other things being equal, a downlink would be capable of generating more key bits over time~\cite{BMHHEHKHDGLJ13}, but an uplink has the advantage of simpler design of the satellite payload (implying reduced risk and cost), and the ability to utilize different source types by exchanging them on the ground. In either case, with QKD states encoded in photon polarization, the key generation rate is fundamentally limited by the total number of photons detected at the receiver.

A major contributing factor to losses experienced over such an optical link is atmospheric turbulence, wherein air pockets of different temperatures lead to varying refractive indices in the transverse and longitudinal modes of the beam path~\cite{VVM19}. This creates atmospheric wavefront errors, manifesting in transverse and temporal intensity fluctuations (scintillation), beam wander, and beam broadening in the far field. This is particularly impactful for the uplink configuration, where the atmospheric wavefront error is induced primarily near the start of the beam propagation, within the first \SI{{\sim}20}{\km} of atmosphere, and exacerbated by the remaining distance to the satellite receiver.

Adaptive optics (AO) utilizes sensors and actuating elements to correct phase errors introduced by atmospheric turbulence~\cite{T10}. Various levels of AO correction can be applied to the optical beam---the simplest is correcting for beam wander as it leaves the transmitter, which corresponds to a tip/tilt correction of the beam, such as would be performed by a conventional closed-loop fine-pointing system. Higher-order corrections can be made by manipulating the phase of the wavefront prior to propagation through the atmosphere. Such approaches are used extensively in astronomical observation~\cite{D08}, optometry~\cite{R11}, and have also been studied for optical communications~\cite{AYW98,PVV16,CLR18}. 

In the context of imaging, AO is used to enhance resolution by compensating for medium-induced aberrations. By contrast, QKD uses polarization analyzers typically coupled to single-pixel (bucket) detectors, so performance is not limited by imaging resolution. In the context of classical communications, where high data bandwidth is desired, AO is employed to minimize scintillation and drop-outs which necessitate overhead owing to error-correction and re-transmission events. There, link stability is the primary concern, with received power being secondary. For QKD (and some other quantum protocols, e.g., Bell tests~\cite{B64}), the total number of measured photons is of greater importance than fluctuations over short time scales, due to the way each photon independently contributes to the protocol~\cite{BMHHEHKHDGLJ13}. As a consequence, we wish to utilize AO to focus the beam more tightly at the receiver in order to increase the total number of photons received within a satellite pass, equivalent to the long-term-averaged power, even though doing so could increase the short-term power variance in the form of bursts and dropouts.

We study the effect of the atmosphere on the long-term-averaged received power to determine how large this effect in itself may be, and model an AO system to determine whether (and by how much) AO may improve optical signal collection at a satellite-based QKD receiver. We do not consider the spectrum of signal power fluctuations (which is greater in uplink due to the motion of the satellite~\cite{TTT11} with respect to the atmosphere), as we assume short-term fluctuations at the receiver remain below the point of detector saturation in the apparatus. As long as the detectors are not operating at saturation, the intensity fluctuations will not negatively impact the key-generation rate (and, when coupled with signal-to-noise filtering, could even provide further performance advantages~\cite{EHMBLWT12}).

Satellite-to-ground quantum links have been studied previously~\cite{GSF16,GFS17,VVM19}, but we focus specifically on the uplink scenario. We model four representative cases of atmospheric conditions relating to ground station locations, with optical links to a satellite receiver of various sizes. Our results show that the potential gains of using adaptive optics with a low-Earth orbit (LEO) and a geostationary (GEO) satellite are modest because of anisoplanatism due to, respectively, fast apparent motion and point-ahead. Correction with a perfectly offset laser guide star could help increase the benefit of AO in these scenarios. In this context, selection of ground station location is the most significant factor determining the optical power that can be captured for use in QKD.

While drafting this manuscript we became aware of Ref.~\cite{OG19}, which also examines the quantum LEO uplink case for AO and makes similar conclusions, there based on a numerical wave optics simulation. Assessment of AO system imperfections in Ref.~\cite{OG19} is done by adjusting the Strehl ratio, without analysis supporting the achievable Strehl ratio. The analysis we present employs an analytical Kolmogorov turbulence model (see Ref.~\cite{P17} for additional details) to derive the impact of atmospheric turbulence on the uplink. We present a comprehensive description of our model and of the wavefront error terms considered, and our performance estimates are based on realistic assumptions of the AO system limitations, while we also consider the impact of ground site selection on AO performance, and examine both the LEO and GEO cases.

\section{Optical Model}\label{sec:opticalModel}

The key criterion for QKD is successful transfer of photonic optical states with high probability, as measured by time-averaged collected power. This directly corresponds to the link efficiency, $\epsilon$, defined as the ratio of the received power, $P_r$, over the transmitted power, $P_t$. Expressed in \si{\dB}, with no AO correction, the link efficiency can be computed from the long-term (time-averaged) beam width (spot radius) at the satellite, $w_\text{LT}$, as~\cite{JWL96}
\begin{align}\label{eqn:efficiency}
\epsilon&=10\log_{10}\left(\frac{P_r}{P_t}\right) \nonumber\\
&=10\log_{10}\left(\eta_r\eta_t\eta^{\sec\psi}_{0}\frac{D^{2}_{r}}{2 w^{2}_\text{LT}}I_\text{LT}\right).
\end{align}
Here, $\eta_r$ is the receiver optical transmittance, $\eta_t$ is the transmitter optical transmittance, $\eta_0$ is the atmosphere optical transmittance at zenith, $\psi$ is the angle of observation from zenith, $D_r$ is the receiver aperture diameter, and $I_\text{LT}\leq1$ (with equality in the ideal case) quantifies the effect of residual beam wander.

The width of an optical beam launched from the ground telescope is affected by diffraction induced by the launch telescope aperture, and by phase error induced by the atmospheric turbulence, which evolve into phase and amplitude errors in the far field. The atmospheric turbulence strength can be quantified by the Fried parameter, or atmospheric turbulence coherence length, $r_0$, which depends on the atmospheric structure constant, $C_n^2(h)$ (for a given altitude $h$), and the air mass that the observer is looking through (which depends on zenith angle, $\psi$). For a spherical wave~\cite{F65},
\begin{equation}
\label{eqn:friedParameter}
r_0=\left[0.423 k^2 \sec\psi \int_{0}^{H} C^{2}_{n}(h) \left(1 - \frac{h}{H}\right)^{5/3}\mathrm{d}h\right]^{-3/5},
\end{equation}
where $H$ is the satellite orbit altitude and $k$ is the wavenumber of the optical beam.

We consider the generalized Hufnagel-Valley (HV) atmospheric structure model~\cite{H74,H98},
\begin{equation}
\begin{split}
C^{2}_{n}(h) = &~A\exp\left(-\frac{h}{H_A}\right)+B\exp\left(-\frac{h}{H_B}\right)\\&+~Ch^{10}\exp\left(-\frac{h}{H_C}\right),
\end{split}
\end{equation}
where $A$ is the coefficient for the surface or boundary layer turbulence strength, $H_A$ is the height for its $1/e$ decay, $B$ and $H_B$ are the equivalent for the turbulence in the troposphere (up to \SI{10}{\km}), and $C$ and $H_C$ are for the turbulence peak at the tropopause (at about \SI{10}{\km}). Further parameters can be included for isolated turbulence layers, but we omit these. This model is used to generate turbulence profiles representing a sea-level site (HV~5-7), an average site (HV~10-10), an excellent site (HV~15-12), and Tenerife~\cite{CGMV04}. The values for $H_A$, $H_B$, and $H_C$ are \SI{100}{\meter}, \SI{1500}{\meter}, and \SI{1000}{\meter} respectively for all four considered models, with the remaining parameters shown in Table~\ref{tab:TurbParameters}.

\begin{table}[t]

	\caption{Turbulence parameters for the generalized HV models of each of the four representative conditions studied~\cite{H74,H98}. $A$ is the coefficient for the surface or boundary layer turbulence strength, $B$ is the equivalent for the turbulence in the troposphere (up to \SI{10}{\km}), and $C$ is for the turbulence peak at the tropopause (at about \SI{10}{\km}). Lower values in these parameters would signify lower turbulence strength. The Tenerife model~\cite{CGMV04} has a turbulence profile between the HV 5-7 (sea-level site) and HV 10-10 (average astronomical site).}
	
\begin{small}
\begin{tabularx}{\columnwidth}{l l l l}

Profile & \multicolumn{3}{l}{Generalized HV model parameters}\\
& $A$ [\si{\meter^{-2/3}}]  & $B$ [\si{\meter^{-2/3}}]  & $C$ [\si{\meter^{-32/3}}]  \\
\hline
HV 5-7   & \num{17e-15}    & \num{27e-17}  & \num{3.59e-53}  \\
HV 10-10 & \num{4.5e-15}   & \num{9e-17}   & \num{2.0e-53}   \\
HV 15-12 & \num{2.0e-15}   & \num{7e-17}   & \num{1.54e-53}  \\
Tenerife & \num{9.42e-15}  & \num{27e-17}  & \num{2.50e-53}  \\

\end{tabularx}
\end{small}
\label{tab:TurbParameters}
\end{table}

\subsection{Closed-loop correction of beam wander}\label{sec:opticalModel_wander}

For an initially Gaussian beam, the long-term $1/e^2$ Gaussian beam width (spot radius) $w_\text{LT}$, when it reaches the satellite at a distance $L$ from the transmitter, is computed by convolving the diffraction-limited width $w_\text{diff}(z) = w_0\sqrt{1 + (z/z_0)^2}$ with the phase-error beam widening from the atmospheric turbulence. Here, $w_0$ is the beam waist and $z_0$ is the Rayleigh distance. This gives~\cite{F75}
\begin{equation}
\label{eqn:ltGaussianWidth}
w_\text{LT}(z=L) = \sqrt{w^{2}_{0}\left(1+\frac{L^2}{z_0^2}\right)+2\left(\frac{4.2L}{kr_0}\right)^2}.
\end{equation}
We neglect the effect of the launch telescope aperture clipping the edge of the Gaussian beam---in typical scenarios, this effect is small compared to other contributions.

Suppose that the transmitter is equipped with a fine tracking system which corrects the beam launch direction based on closed-loop measurement of a beacon laser reference transmitted from the satellite. In this case, atmospheric tilt effects within the bandwidth of this system will be compensated, and the long-term beam width, $w_\text{LT}$, can be modeled as a short-term beam width, $w_\text{ST}$, that is broadened by the residual beam wander. This short-term beam width, which we will utilize later, is given by~\cite{Y73}
\begin{equation}
\begin{split}
&w_\text{ST}(z=L) =\\ &\Biggl[w^{2}_{0}\left(1+\frac{L^2}{z_0^2}\right)+ 2\left(\frac{4.2L}{kr_0}\left[1-0.26\left(\frac{r_0}{w_0}\right)^{1/3}\right]\right)^2\Biggr]^{1/2}.
\end{split}
\end{equation}

$I_\text{LT}$ is computed by assuming that the two-dimensional residual beam wander has Gaussian statistics with standard deviations that are added (in quadrature) to the long-term beam width~\cite{JWL96}. We take the mean, calculated generally as $\langle I \rangle = \beta / (\beta + 1)$ where $\beta = (\Theta / \sigma)^2 / 8$. Here, $\Theta \approx W/L$ is the full angle beam divergence for a beam width, $W$ (diameter), at the satellite (for $I_\text{LT}$, $W = \sqrt{2}w_\text{LT}$) and $\sigma$ is the one-dimensional residual beam wander standard deviation.

The residual beam wander, $\sigma$, mainly depends on four error sources: the limited signal-to-noise ratio of the beacon sensor ($\sigma_\text{SNR}$), the closed-loop feedback delay of the fine-pointing system ($\sigma_\text{TFD}$), centroid anisoplanatism ($\sigma_\text{CA}$), and tilt anisoplanatism ($\sigma_\text{TA}$). Note that our model assumes the telescopes are physically pointing exactly at each other (with appropriate point-ahead).

$\sigma_\text{SNR}$: Typically achievable signal-to-noise ratios of commercially available position sensitive devices (PSDs) lead to $\sigma_\text{SNR} < \SI{0.15}{\micro\radian}$ with bright sources. $\sigma_\text{SNR} = \SI{0.15}{\micro\radian}$ is assumed in our model---see, e.g., Ref.~\cite{FS}. This error source typically has a small contribution to the overall error.

$\sigma_\text{TFD}$: Tilt feedback delay error is caused by the atmospheric-induced tilt evolving from the time it is read by the sensor to the moment the correction is applied. For a fine-pointing system with a closed-loop correction bandwidth $f_\text{c}$, it is calculated as~\cite{T94}
\begin{equation}
\label{eqn:tfd}
\sigma_\text{TFD}=\frac{f_\text{T}}{f_\text{c}} \frac{\lambda}{D_t},
\end{equation}
where $\lambda$ is the optical wavelength, $D_t$ is the ground transmitter telescope diameter, and $f_\text{T}$ is the tracking frequency, defined as the frequency at which the one-sigma $\sigma_{\text{TFD}}$ is equal to the diffraction angle $\lambda/D_t$. Given $C^{2}_{n}(h)$ and wind speed profile $v_w(h)$, the tracking frequency is
\begin{equation}
f_\text{T}=0.331D^{-1/6}_{t}\lambda^{-1}\left[\sec\psi\int^{H}_{0}C^{2}_{n}(h)v^{2}_{w}(h) \mathrm{d}h\right]^{1/2}.
\end{equation}
The wind speed profile is based on a Bufton wind model~\cite{B73,H98,MJC10,TTT11},
\begin{equation}
\label{eqn:buftonWind}
v_w(h) = v_g + v_t \exp\left[-\left(\frac{h - h_\text{peak}}{h_\text{scale}}\right)^2\right] \mathbin{+} h \dot\psi,
\end{equation}
where $v_g$ is the ground wind speed (\SI{5}{\m/\s}), $v_t$ is the high-altitude wind speed (\SI{20}{\m/\s}), $h_\text{peak}$ is the altitude of the peak (\SI{9.4}{\km}), $h_\text{scale}$ is the scale height (\SI{4.8}{\km}), and $\dot\psi$ is the angular velocity of the satellite apparent to the ground station. To calculate $\dot\psi$, we consider a simplified model of an object in circular orbit around a spherical Earth, resulting in a constant Earth-centred angular velocity. From this we derive $\dot\psi$ for a ground station at Earth's surface targetting a satellite at \SI{600}{\km} altitude.

$\sigma_\text{CA}$: Higher order wavefront errors induced by turbulence eddies smaller than the aperture of the transmitter telescope change the point spread function (PSF) shape incident on the PSD that leads to centroid estimation errors, or \emph{centroid anisoplanatism}. The one-dimensional standard deviation for this is given by~\cite{T94}
\begin{equation}
\sigma_\text{CA} = 5.51\times10^{-2}\left(\frac{\lambda}{D_t}\right)\left(\frac{D_t}{r_0}\right)^{5/6}.
\end{equation}
This term is mostly dependent on the turbulence strength as determined by the Fried parameter, and is \SI{{\sim}0.4}{\micro\radian} for an HV 5-7 model for transmissions at zenith.

$\sigma_\text{TA}$: The finite speed of light coupled with the distance and motion of the satellite requires the ground station to transmit optical beams ahead of the satellite's apparent position at any given time to ensure they are caught by the satellite receiver. This implies that, at the time of measurement and correction, the satellite's downlink beacon will have taken a different path through the atmosphere than the transmitted beam will take back to the satellite, thereby leading to \emph{tilt anisoplanatic} error. Originally derived in Ref.~\cite{F76} and following Ref.~\cite{OMGB92}, this error is
\begin{equation}
\label{eqn:ta}
 \sigma_\text{TA} = 6.14 D_t^{-1/6} \left[\sec\psi \int_0^H C^2_n(h) f_\Delta(h) \mathrm{d}h \right]^{1/2},
\end{equation}
where $f_{\Delta}$ is a weighting function for a circular aperture given by
\begin{equation}
 \begin{split}
 f_\Delta(h) &= \int_0^{2\pi} \int_0^1 \Bigl[ \frac{1}{2} (u^2 + 2us\cos w + s^2)^{5/6} \\&+ \frac{1}{2}(u^2 - 2us\cos w + s^2)^{5/6} - u^{5/3} - s^{5/3} \Bigr] \\
 &\qquad\qquad \times u [ \cos^{-1} u - (3u - 2u^3)\sqrt{1 - u^2} ] \mathrm{d}u \mathrm{d}w,
 \end{split}
\end{equation}
and
\begin{equation}
 s = \frac{\delta_\text{tilt} h \sec \psi}{D_t}.
\end{equation}

$\delta_\text{tilt}$ quantifies the change in tilt angle between the incoming and outgoing beams---the point-ahead angle---and depends on $\dot\psi$. From our orbit model, $\delta_\text{tilt}$ is then determined for a surface ground station, with the object's orbit passing zenith, using the round-trip time necessary for light to propagate the length of the incoming and outgoing beam paths. For a satellite orbiting at \SI{600}{\km}, $\delta_\text{tilt}$ is \SI{50}{\micro\radian} at zenith.

\subsection{Adaptive optics correction of wavefront error}\label{sec:opticalModel_adaptive}

We now introduce to our model adaptive optics to correct higher-order wavefront aberrations. The received beam at the satellite produced by such a system is modelled by a diffraction-limited core surrounded by a seeing-limited halo~\cite{T10}. The link efficiency equation can be recast as
\begin{equation}
\label{eqn:aoEfficiency}
\epsilon =  10\log_{10}\left(\eta_r\eta_t\eta^{\sec\psi}_{0}\frac{D^{2}_{r}}{2}\left[\frac{I_\text{diff}S}{w^{2}_\text{diff}} 
 +\frac{I_\text{ST}(1-S)}{w^{2}_\text{ST}}\right]\right),
\end{equation}
where $w_\text{diff}$ is the diffraction-limited beam width, and $I_\text{diff}$ (equalling $\langle I \rangle$ with $w=w_\text{diff}$) and $I_\text{ST}$ (equalling $\langle I \rangle$ with $w = w_\text{ST}$) quantify the energy loss due to residual beam wander for the diffraction-limited core and the short-term seeing-limited halo, respectively. 
Note that as $\sigma \rightarrow 0$, $I_\text{LT}$, $I_\text{diff}$, and $I_\text{ST}$ all approach 1---for convenience, we notate this limiting case as $I=1$. 
The Strehl ratio $S$ is defined as the fraction of optical power that is in the diffraction-limited core compared to a perfectly corrected system. Better AO wavefront correction leads to a higher Strehl ratio. 

For a given root-mean-squared (RMS) wavefront error in radians (for which $2\pi$ radians equates to an error of $\lambda$), it is evaluated from the Mahajan equation~\cite{M83}, $S \approx \exp ({-\zeta^2})$. Note that we label the wavefront error $\zeta$, in contrast to common treatments, in order to avoid confusion with the residual beam wander $\sigma$ and its contributing terms.

We consider three sources of error in our model of an AO system, the standard deviations of which are added in quadrature to determine $S$: the AO feedback delay error ($\zeta_\text{AFD}$), the spatial fitting error ($\zeta_\text{fit}$), and the phase anisoplanatic error ($\zeta_\text{PA}$),
\begin{equation}
\zeta^2=\zeta_\text{AFD}^2+\zeta_\text{fit}^2+\zeta_\text{PA}^2.
\end{equation}

$\zeta_\text{AFD}$: The AO feedback delay error is similar in origin to the tilt feedback delay error ($\sigma_\text{TFD}$), being caused by the atmospheric turbulence evolving between the time the wavefront error is measured and the time it is corrected. In the case of higher-order aberrations, the tracking frequency is replaced by the Greenwood frequency~\cite{T10,G77},
\begin{equation}
f_\text{G}=2.31\lambda^{-6/5}\left[\sec\psi\int^{H}_{0}C^{2}_{n}(h) v^{5/3}_{w}(h) \mathrm{d} h\right]^{3/5}.
\end{equation}
The associated wavefront error term is~\cite{T10}
\begin{equation}
\zeta_\text{AFD}=\left(\frac{f_\text{G}}{f_\text{c}}\right)^{5/6}.
\end{equation}
This is mainly dependent on the wavelength and the turbulence strength. 

$\zeta_\text{fit}$: The spatial fitting error is caused by the limited degrees of freedom of the wavefront corrector. Assuming perfect control based on the Zernike polynomials~\cite{T10,Z34}, equations defining the residual wavefront error after correction of $J$ Zernike polynomials under Kolmogorov turbulence can be found in Table~IV of Ref.~\cite{N76}, for $J$ up to 21. For $J>10$, these can be approximated with~\cite{N76}
\begin{equation}
\zeta_\text{fit}^2 = 0.2944 J^{-\sqrt{3}/2}\left(\frac{D_t}{r_0}\right)^{5/3}.
\end{equation}

$\zeta_\text{PA}$: The phase anisoplanatic error is given by~\cite{T10}
\begin{equation}
\zeta_\text{PA}=\left(\frac{\theta}{\theta_0}\right)^{5/6}.
\end{equation}
Here, $\theta$ is the angle between the reference beam and the corrected beam (usually, $\theta = \delta_\text{tilt}$), and the \emph{isoplanatic angle} $\theta_0$ is the angle between the object and the reference beam at which the wavefront variance is \SI{1}{\square\radian}, which evaluates to
\begin{equation}
\label{eqn:isoplanatic}
\theta_0 = \left[2.91k^2(\sec\psi)^{8/3}\int^{H}_{0}C^{2}_{n}(h)h^{5/3} \mathrm{d}h\right]^{-3/5}.
\end{equation}

\section{Application to a particular Earth-station-to-satellite scheme}
\label{sec:analysis}

The model described in the previous sections is used to evaluate the impact of atmospheric turbulence on a ground-to-satellite link and the potential improvement achievable with an AO system. The base parameters of the physical system considered are given in Table~\ref{tab:SatParameters}. These are the parameters we use in our model to produce the results given below, unless indicated otherwise.

\begin{table}[t]

\centering
\begin{small}
\begin{tabularx}{\columnwidth}{l l  l }
Parameter & Symbol & Value \\
\hline
Satellite altitude [\si{\kilo\metre}] & $H$ & 600 \\
Receiver aperture diameter [\si{\m}] & $D_r$ & 0.4 \\ 
Receiver optical transmittance & $\eta_r$ & 0.5 \\
QKD signal wavelength [\si{\micro\meter}] & $\lambda$ & 0.785 \\
Transmitter aperture diameter [\si{\meter}] & $D_t$ & 0.50 \\
Transmitter optical transmittance & $\eta_t$ & 0.5 \\
Optical transmittance at zenith & $\eta_0$ & 0.8 \\
Atmosphere model &  & HV 5-7 \\
High-altitude wind speed [\si{\meter/\second}] & $v_t$ & 20 \\
Correction bandwidth [\si{\Hz}] & $f_\text{c}$ & 200 \\
Zernike polynomials corrected & $J$ & 45 \\

\end{tabularx}
 \end{small}
\label{tab:SatParameters}
\caption{Summary of the ground-to-satellite link baseline parameters for the simulations. These parameters are used in the simulation unless otherwise stated.}
\end{table}

For the \SI{0.5}{\meter} transmitter diameter, we chose to correct Zernike polynomials up to the 8th order in our simulation, as we observed minimal increase in the Strehl ratio for higher order compensation. This implies that the first 45 Zernike terms are corrected. Such can be achieved, for example, by using a Shack-Hartmann or a pyramidal wavefront sensor having 8 sub-apertures on the pupil diameter, associated with a deformable mirror having 9 linear actuators on the pupil diameter. The additional actuator is required because the wavefront sensor measures a wavefront slope which is compensated using one actuator at each edge of the sub-aperture in one dimension, and one actuator at each corner of the sub-aperture in two dimensions. (This arrangement of actuators relative to sub-apertures is typically referred to as the Fried geometry.)

\begin{figure*}[tbp]
\centering
\includegraphics[width=\columnwidth]{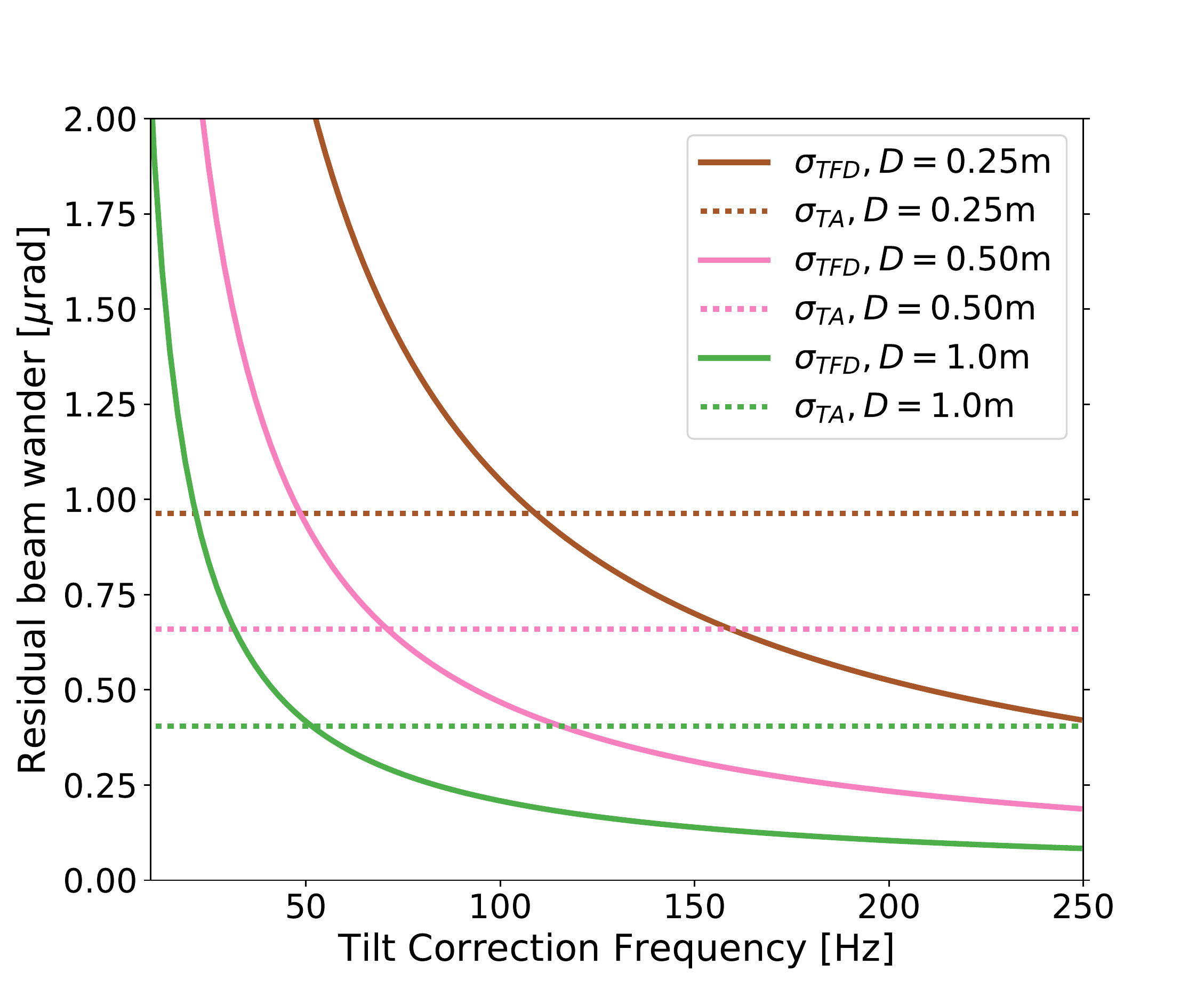}
\includegraphics[width=\columnwidth]{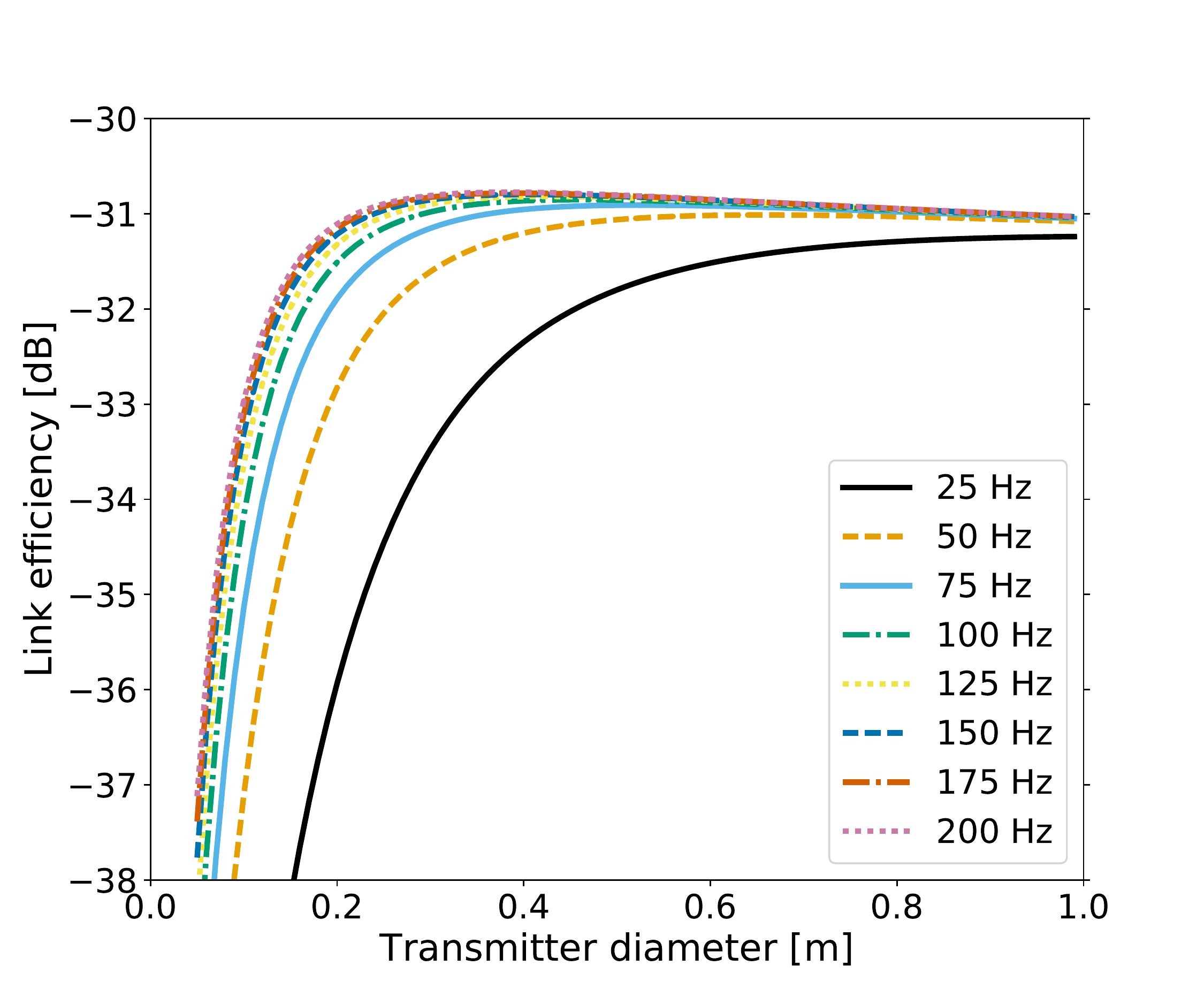}
\caption{(Left) $\sigma_\text{TA}$ (Eq.~\ref{eqn:ta}) and $\sigma_\text{TFD}$ (Eq.~\ref{eqn:tfd}) contributions to residual beam wander as functions of beam wander correction bandwidth for various transmitter diameters. As the diameter is decreased, the system requires a higher beam wander correction bandwidth for $\sigma_\text{TA}$ to remain the dominant error term. (Right) Link efficiency (Eq.~\ref{eqn:efficiency}) as a function of transmitter diameter for various beam wander correction bandwidths.}
\label{fig:freqError}
\end{figure*}

For low correction bandwidths and small transmitter diameters, the dominating tilt error term is the tilt feedback delay ($\sigma_{\text{TFD}}$). Once the correction bandwidth and transmitter diameter are sufficiently increased, the tilt anisoplanatic error ($\sigma_{\text{TA}}$) dominates. This can be seen in Figure~\ref{fig:freqError}. For a \SI{0.5}{\metre} transmitter, the correction frequency beyond which $\sigma_\text{TA}$ dominates is \SI{{\approx}70}{\Hz}. At \SI{{\approx}200}{\Hz}, $\sigma_\text{TFD}$ is \SI{{\sim}35}{\percent} of $\sigma_\text{TA}$. (Bandwidths beyond \SI{{\sim}200}{\Hz} are a considerable technical challenge to implement, as the sampling rate must be greater by a factor of 10 to 20.)

The maximum potential impact of correcting the residual beam wander can be quickly evaluated by assuming a perfect tilt correction. Substituting the long-term beam width in Eq.~\ref{eqn:efficiency} with the short-term value and correcting perfectly for the beam wander by setting $I = 1$ yields a mere \SIrange{1}{3}{\dB} improvement in the link efficiency at zenith for each of the modeled atmospheres and various transmitter aperture diameters (\SIrange{0.20}{1.0}{\meter}). Evidently, beam wander correction is not a path towards any significant gain in performance.

\begin{figure*}[tbph]
	\centering
	\includegraphics[width=0.66\columnwidth]{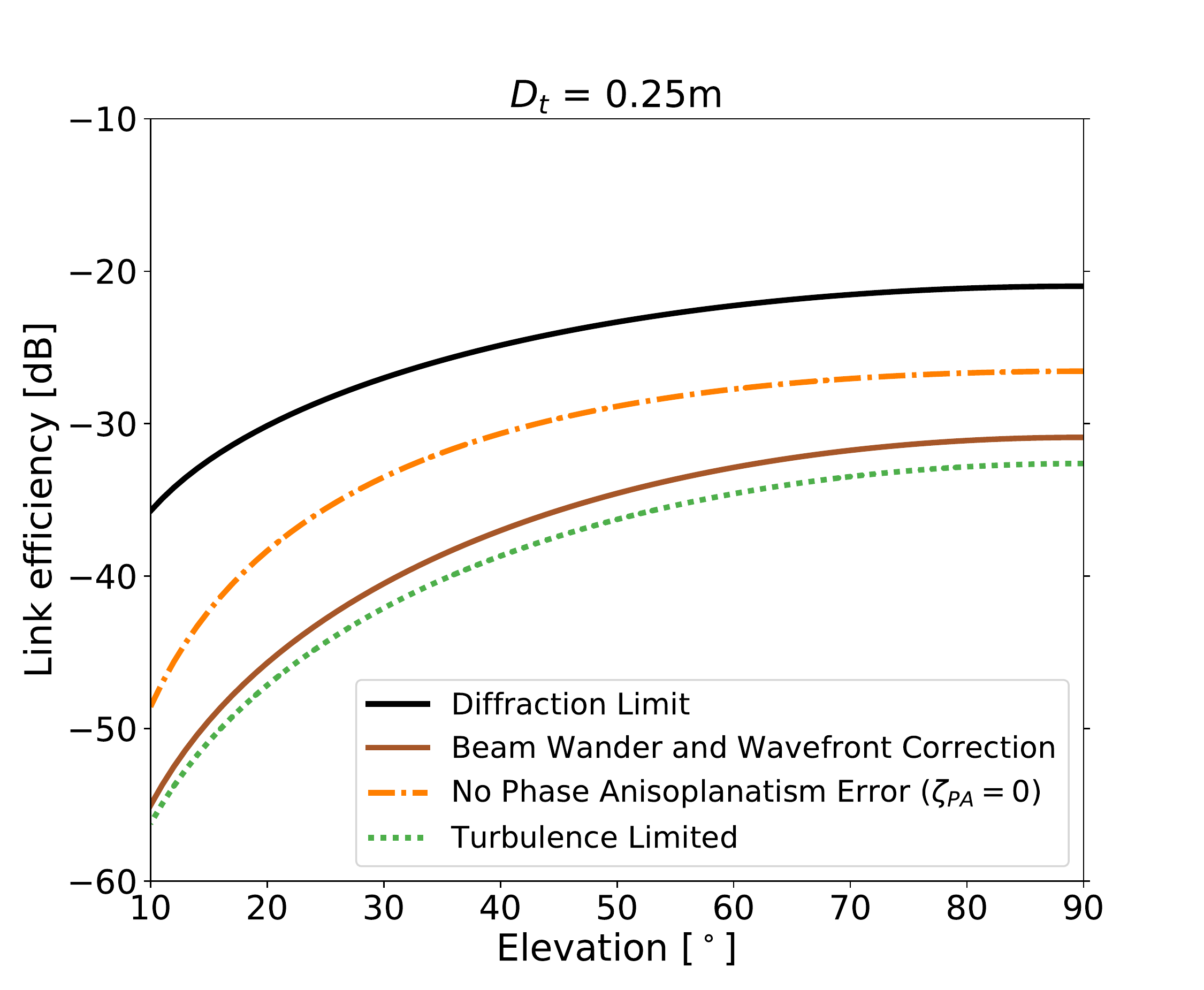}
	\includegraphics[width=0.66\columnwidth]{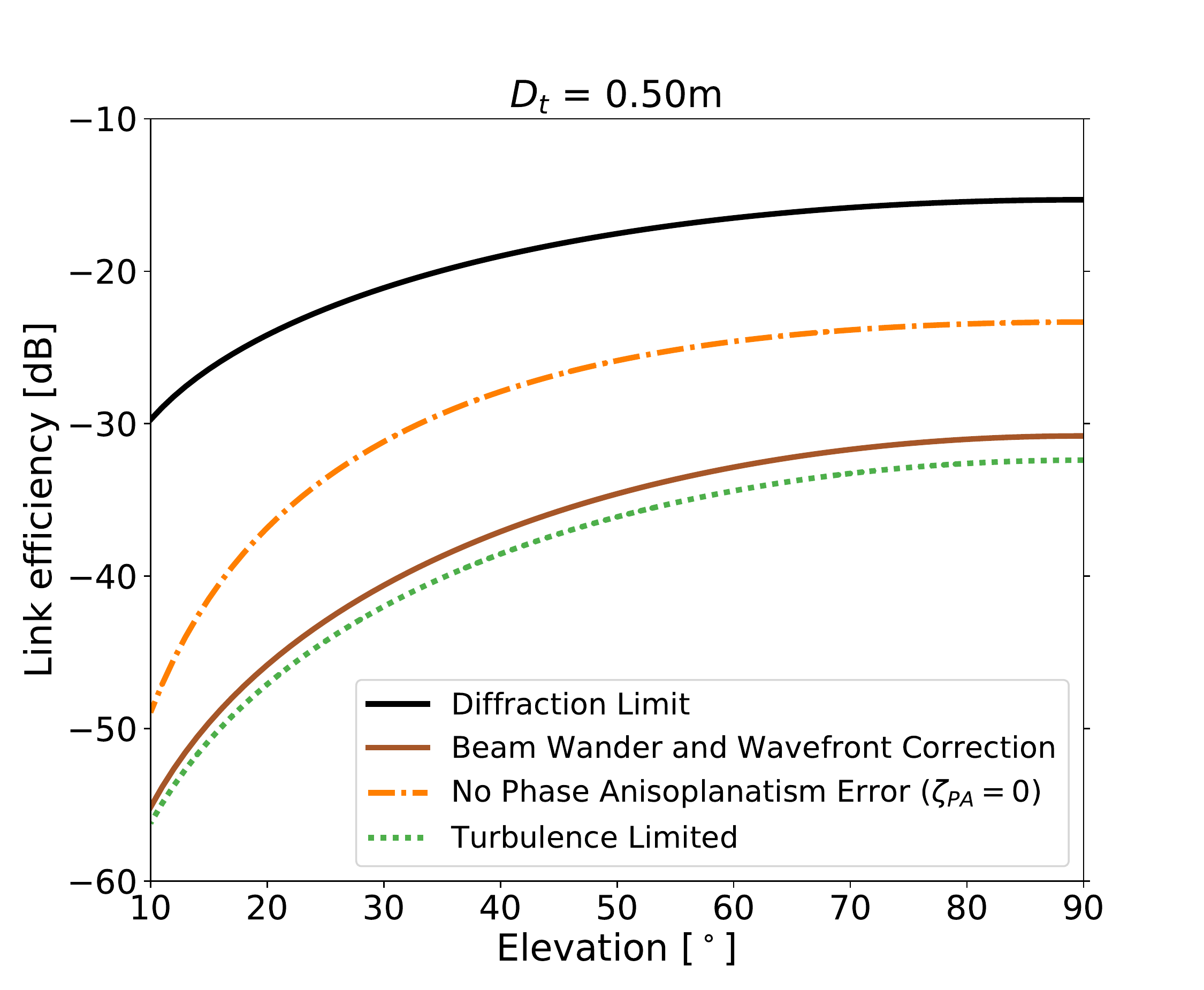}
	\includegraphics[width=0.66\columnwidth]{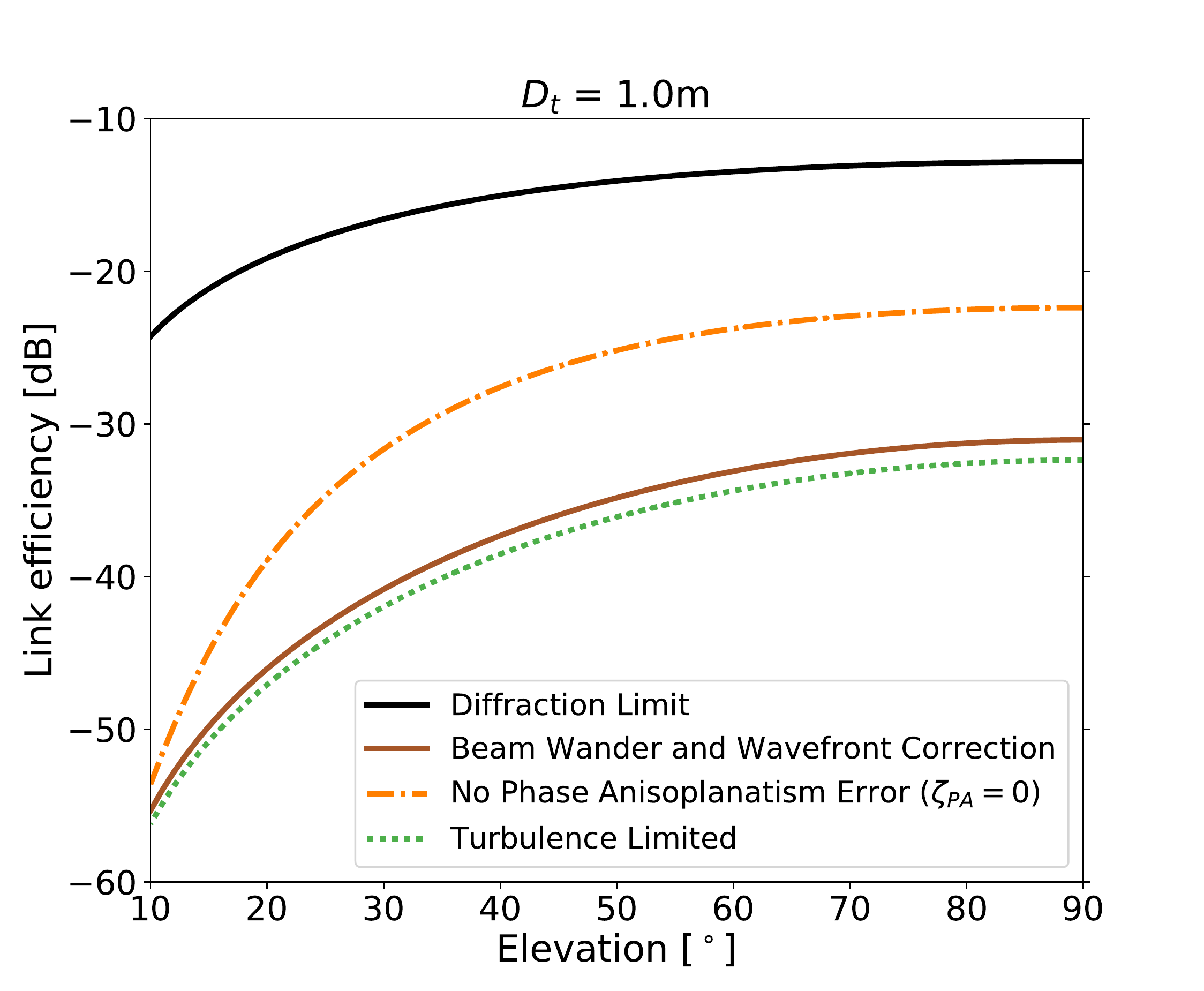}
	\caption{Comparison of the predicted link efficiency (Eq.~\ref{eqn:aoEfficiency}) as a function of elevation angle of the satellite, from horizon apparent to the ground station, with AO for transmitter diameters of \SI{0.25}{\meter} (left), \SI{0.50}{\meter} (middle), and \SI{1.0}{\meter} (right). The solid black curve shows the diffraction-limited efficiency. The dotted green curve, the baseline case, shows a turbulence-limited beam with no corrections. The solid brown curve shows the efficiency for correcting both beam wander and wavefront phase errors. The dash-dotted orange curve represents correcting for beam wander and wavefront phase errors, but assuming that the wavefront phase anisoplanatism error term, $\zeta_\text{PA}$, is zero. The phase anisoplanatism error term dominates the error sources, and if not corrected renders an AO solution only marginally useful.}
	\label{fig:diameterChange}
\end{figure*}

We now consider how the inclusion of AO correction to the wavefront error affects the link performance (Eq.~\ref{eqn:aoEfficiency}) for various launch telescope diameters. The results are shown in Figure~\ref{fig:diameterChange}. Four scenarios are contained in each plot, representing the diffraction-limited beam link efficiency, the baseline case where a beam propagates through a turbulent medium with no corrections applied, the case where beam wander and wavefront phase errors are corrected, and the same case while assuming no wavefront phase anisoplanatism term ($\zeta_\text{PA}=0$).

A larger transmitter aperture produces a smaller diffraction-limited spot at the receiver location. The potential for adaptive optics to improve the link efficiency above the turbulence-limited baseline is then also greater---see the dash-dotted orange lines in Fig.~\ref{fig:diameterChange}. At zenith, the additional gain is as much as \SI{3.2}{\dB} if a \SI{0.5}{\m} transmitter is used, and \SI{4.2}{\dB} for a \SI{1.0}{\m} transmitter, compared to the \SI{0.25}{\m} transmitter efficiency. However, when we include the anisoplanatic error that arises because the downlink and uplink beams do not follow the same paths through the atmosphere, these gains are lost.

\begin{figure*}[tbp]
\centering
\includegraphics[width=\columnwidth]{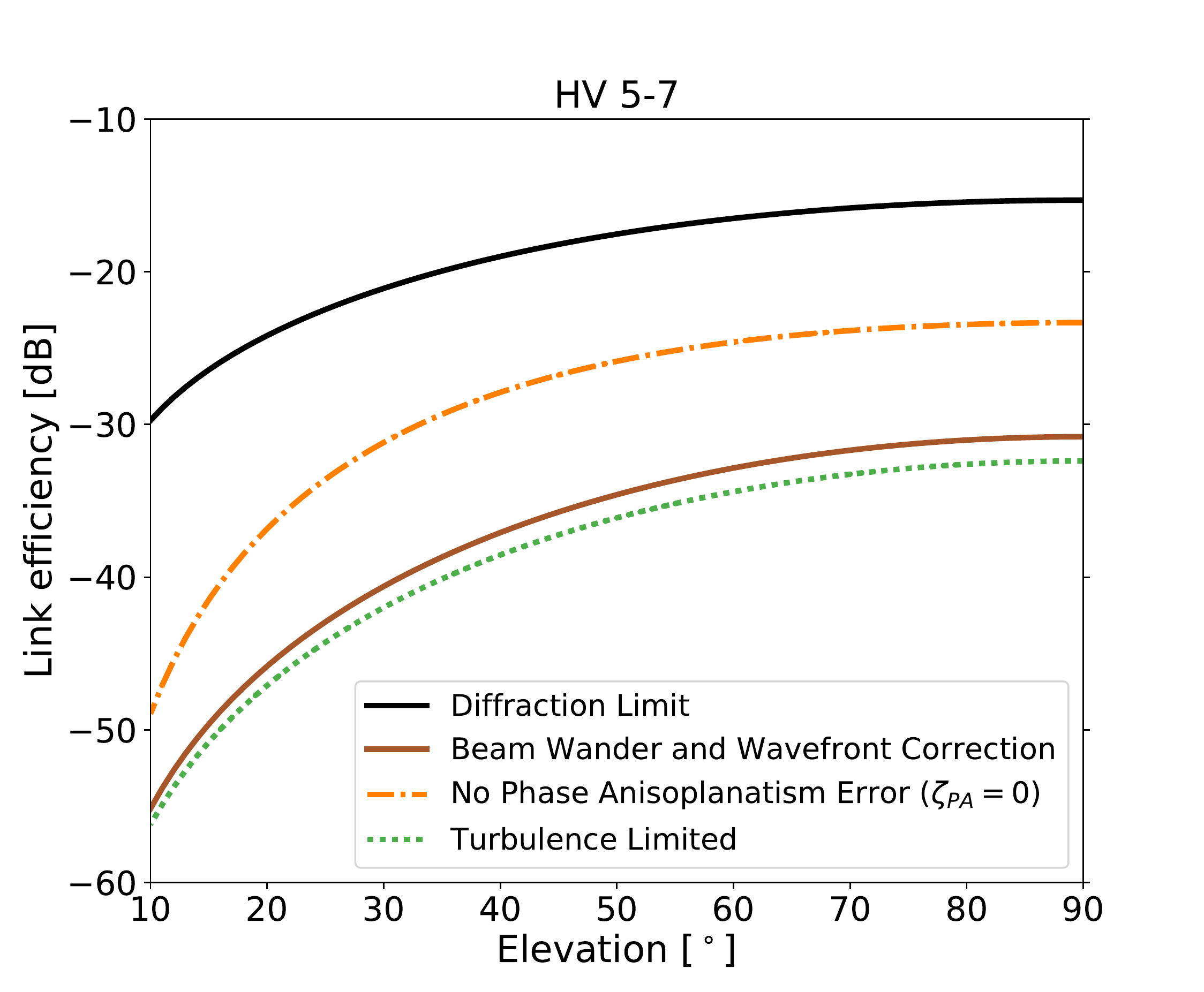}
\includegraphics[width=\columnwidth]{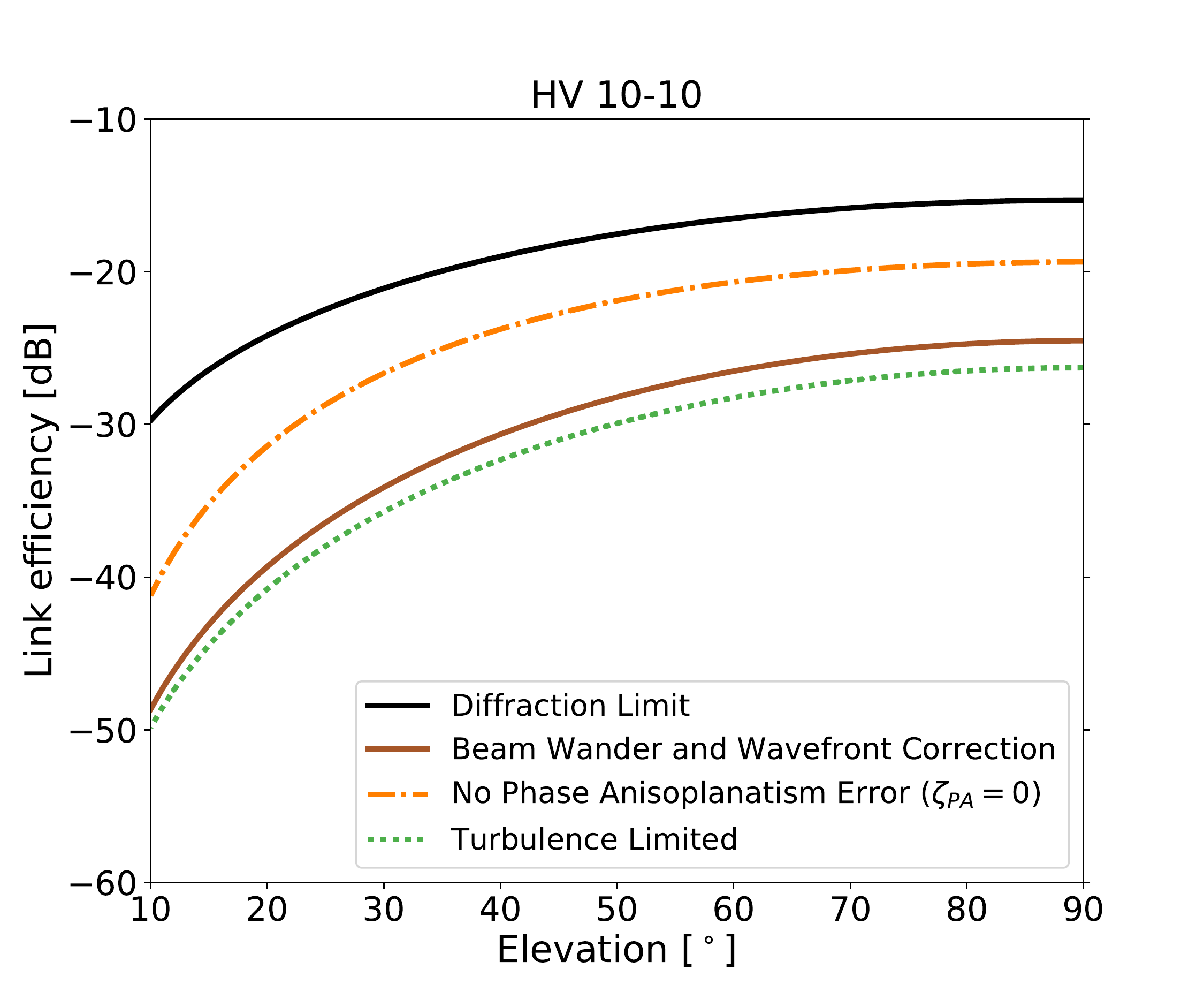}\\
\includegraphics[width=\columnwidth]{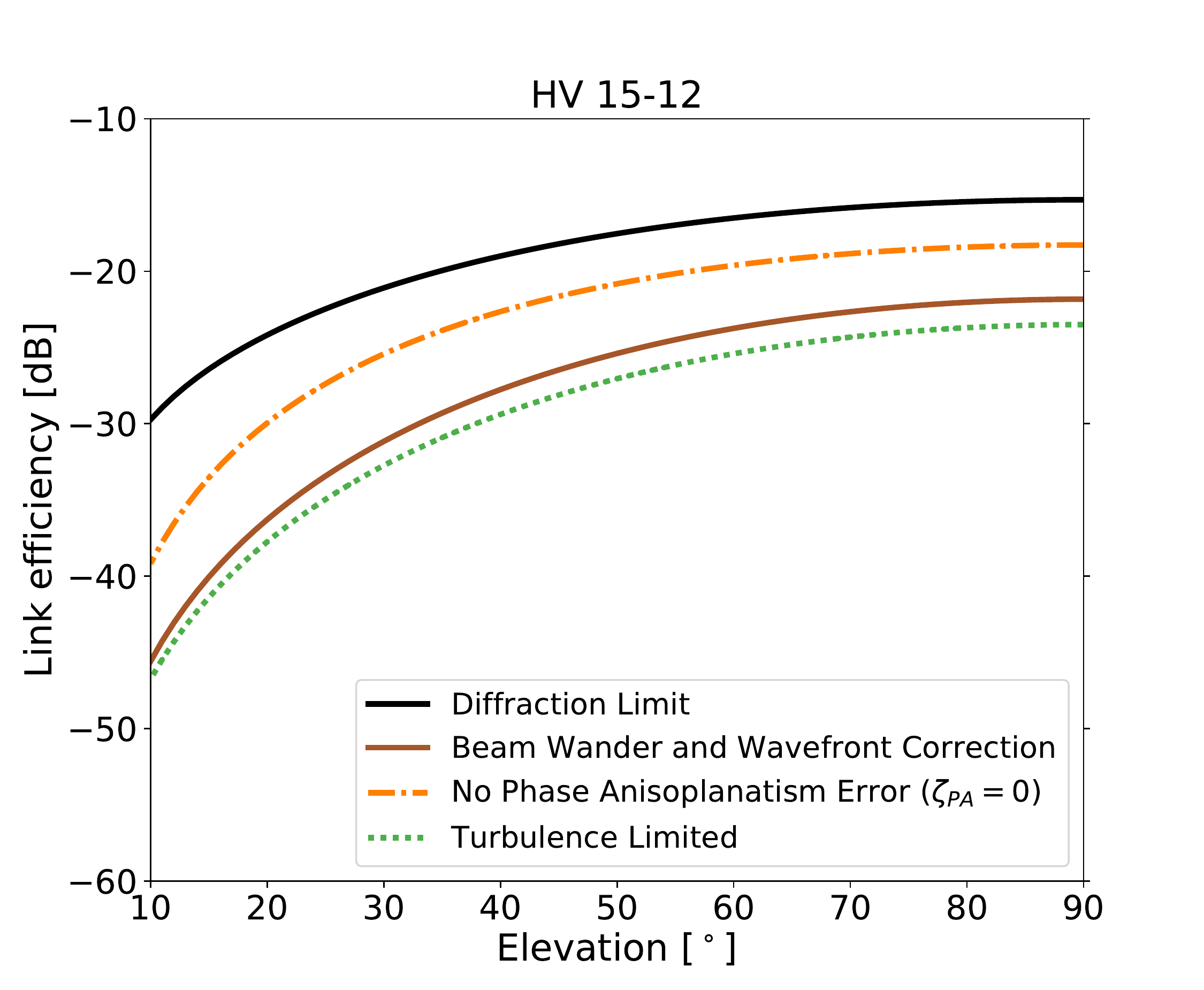}
\includegraphics[width=\columnwidth]{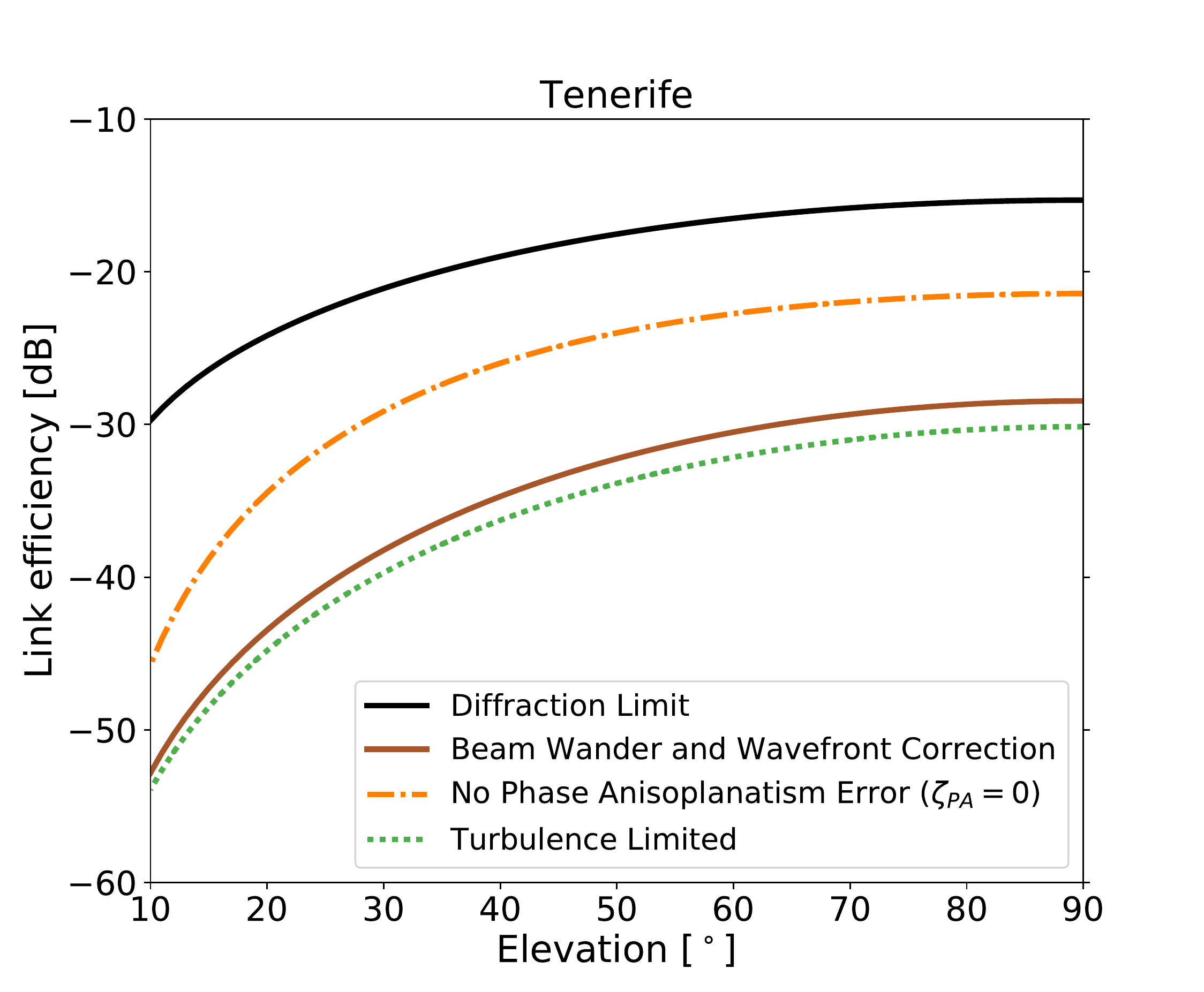}
\caption{Prediction of the efficiency (Eq.~\ref{eqn:aoEfficiency}) as a function of elevation angle with AO for HV 5-7, a typical sea-level site (top-left), HV 10-10, a typical good site (top-right), HV 15-12, an excellent site (bottom-left), and Tenerife, from measured median turbulence strength at Tenerife in the Canary Islands~\cite{CGMV04} (bottom-right). Curves follow definitions in Figure~\ref{fig:diameterChange}.
Adaptive optics demonstrates slightly better performance improvement with stronger turbulence profiles.}
\label{fig:totalCorrection}
\end{figure*}

Figure~\ref{fig:totalCorrection} illustrates the effect of different turbulence strengths, using the parameters given in Table~\ref{tab:SatParameters}. This shows that the turbulence strength can have a significant impact on the link efficiency and, therefore, site selection is a critical factor determining throughput. A gain of approximately \SI{9}{\dB} is found for a good astronomical site (HV 15-12) compared to a site at sea-level (HV 5-7). What small improvement adaptive optics can achieve, however, is outweighed by selection of better sites.

As with the residual beam wander correction, the AO performance is strongly limited by anisoplanatism, to the extent that there is little gain in using an AO system to correct the wavefront phase errors if the responsible phase anisoplanatic error term is not mitigated. This strong contribution of the anisoplantic error is symptomatic of a weak correlation between the turbulence in the downlink and uplink paths. 

For an AO system to be useful, it is critical to reduce the anisoplanatic phase error term. A common and mature approach for doing so in astronomical applications is to use a reference laser guide star (LGS) to sample the turbulence in the proper atmospheric path~\cite{TF90}. This LGS can be generated by exciting atoms in the \SI{90}{\km} altitude sodium layer with a laser, or to use a time-gating camera to observe the Rayleigh backscatter of a pulsed laser at an altitude of typically \SI{18}{\km}. While this LGS mitigates the phase anisoplanatic error ($\theta = 0$, and thus $\zeta_\text{PA} = 0$), a new error term needs to be considered due to the difference in altitude between the satellite and the LGS, which results in a different path taken through the atmosphere by the light emitted by the LGS (and subsequently captured by the telescope) and the quantum signal travelling to the satellite. This is referred to as focal anisoplanatism or cone effect, and the corresponding error term can be expressed as~\cite{T10}
\begin{equation}
\label{eqn:coneError}
\zeta_{\text{cone}}=\left(\frac{D_t}{d_0}\right)^{5/6},
\end{equation}
where
\begin{equation}
\small
\label{eqn:lgsDist}
d_0 = \lambda^{6/5} \left(19.77\sec\psi\int_{0}^{H_\text{LGS}}C_n^2(h)\left(\frac{h}{H_\text{LGS}}\right)^{5/3}\text{d}h\right)^{-3/5},
\end{equation}
and $H_\text{LGS}$ is the altitude of the LGS.

\begin{figure}[tbp]
	\centering
	\includegraphics[width=\columnwidth]{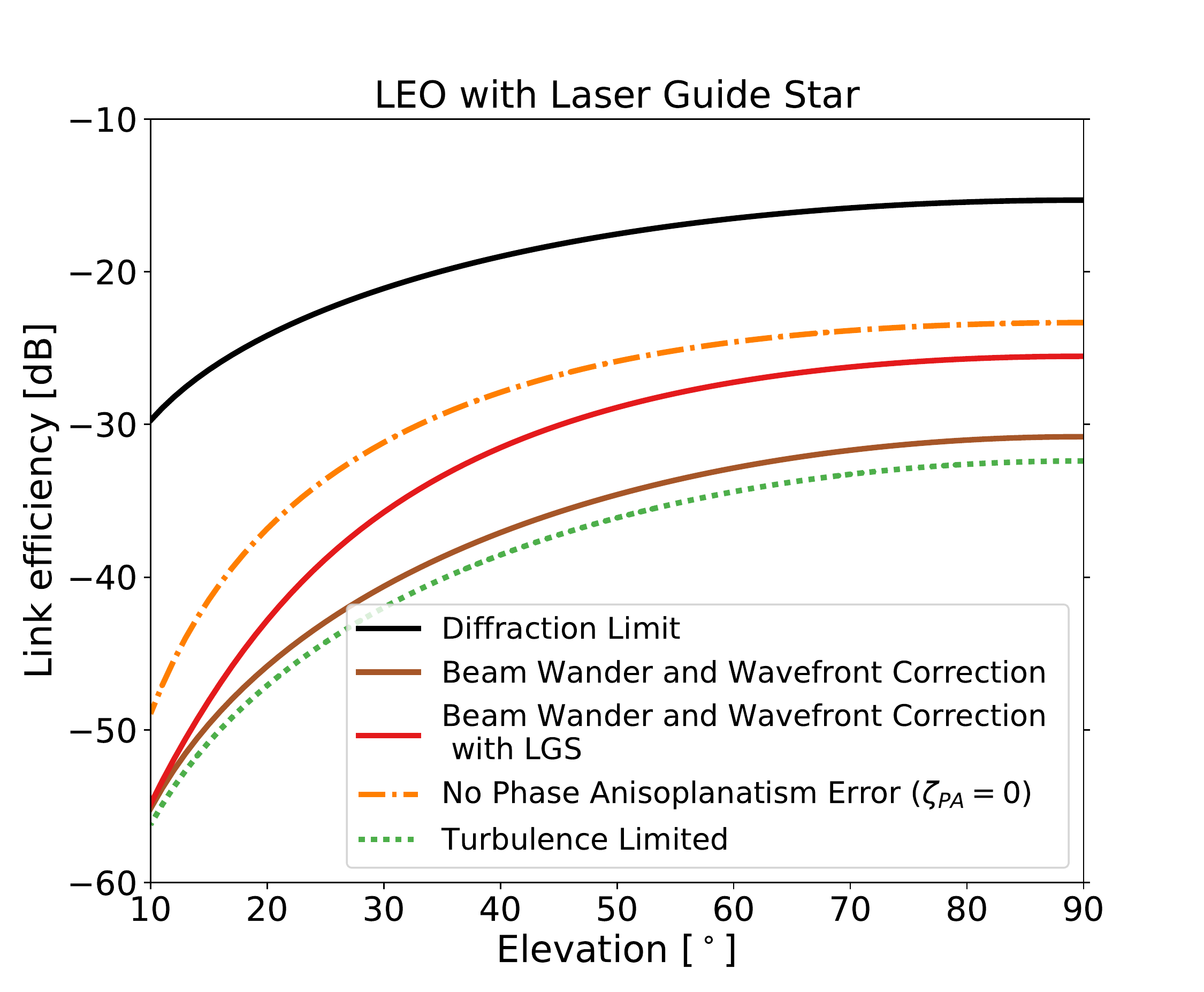}
	\caption{Predicted link efficiency as a function of elevation angle using a laser guide star at \SI{18}{\kilo\meter}. Curves follow definitions in Figure~\ref{fig:diameterChange}, with the addition of the solid red curve showing the efficiency using a laser guide star to correct for the wavefront phase anisoplanatism error. In comparison to AO without LGS, a significant improvement of approximately \SI{6.4}{\dB} in the HV 5-7 atmosphere model is seen at zenith.}
	\label{fig:LGS}
\end{figure}

The result is illustrated in Figure~\ref{fig:LGS}, showing an improvement from using a LGS at \SI{18}{\kilo\meter} (\SI{{\approx}5.3}{\dB} at zenith, compared to no LGS) to an overall \SI{{\approx}6.9}{\dB} efficiency increase at zenith when compared to the baseline. Table~\ref{tab:Strehls} shows example Strehl ratios under these conditions---by using a LGS to reduce the anisoplanatism, the Strehl ratio can be dramatically increased, approaching the values asserted in Ref.~\cite{OG19}. Note the LGS cannot also be used as a reference for beam wander correction (tip-tilt) because the exact LGS location is not well defined due to the laser beam being affected by some beam wander in both its upward propagation, on its way to generate the LGS, and its downward propagation. Therefore, we use the LGS to correct the higher order wavefront corrections ($\zeta$ terms) and use the beacon from the satellite to correct the beam wander ($\sigma$ terms). This is equivalent to using Eq.~\ref{eqn:aoEfficiency} as before, however, replacing $\zeta_\text{PA}$ with $\zeta_\text{cone}$.

\begin{table}[t]
	
	\centering	
	\begin{tabularx}{\columnwidth}{l l  l }
		Atmosphere:& HV 5-7 & HV 15-12\\
		
		Transmitter Diameter &  &\\
		\hline
		\SI{0.25}{\meter}&\num{0.359}&\num{0.681}\\
		\SI{0.5}{\meter}&\num{0.198}&\num{0.555}\\
		\SI{1.0}{\meter}&\num{2.98e-2}&\num{0.289}\\
	\end{tabularx}	
	\label{tab:Strehls}
	\caption{Strehl ratios when using a LGS at \SI{18}{\kilo\meter}. Note that the decrease of Strehl ratio for increasing transmitter diameter is due to the number of corrected Zernike terms being constant, implying increased spacing between the deformable mirror actuators and increased sub-aperture size in relation to $r_0$. }
\end{table}

Geostationary (GEO) satellites orbit the Earth at approximately \SI{35000}{\kilo\metre} and have the same orbital period as the Earth's rotational period. Consequently, the satellite appears stationary in the sky relative to a ground station located anywhere on Earth. For this reason, intuitively, one could expect that anisoplanatism would be near minimal in this case. Also, a geostationary orbit is interesting for communications (including quantum communications) as it provides coverage over up-to half of Earth at any given time, unlike LEO which only provides coverage to a given ground station during limited time windows.

\begin{figure}[tbp]
	\centering
	\includegraphics[width=\columnwidth]{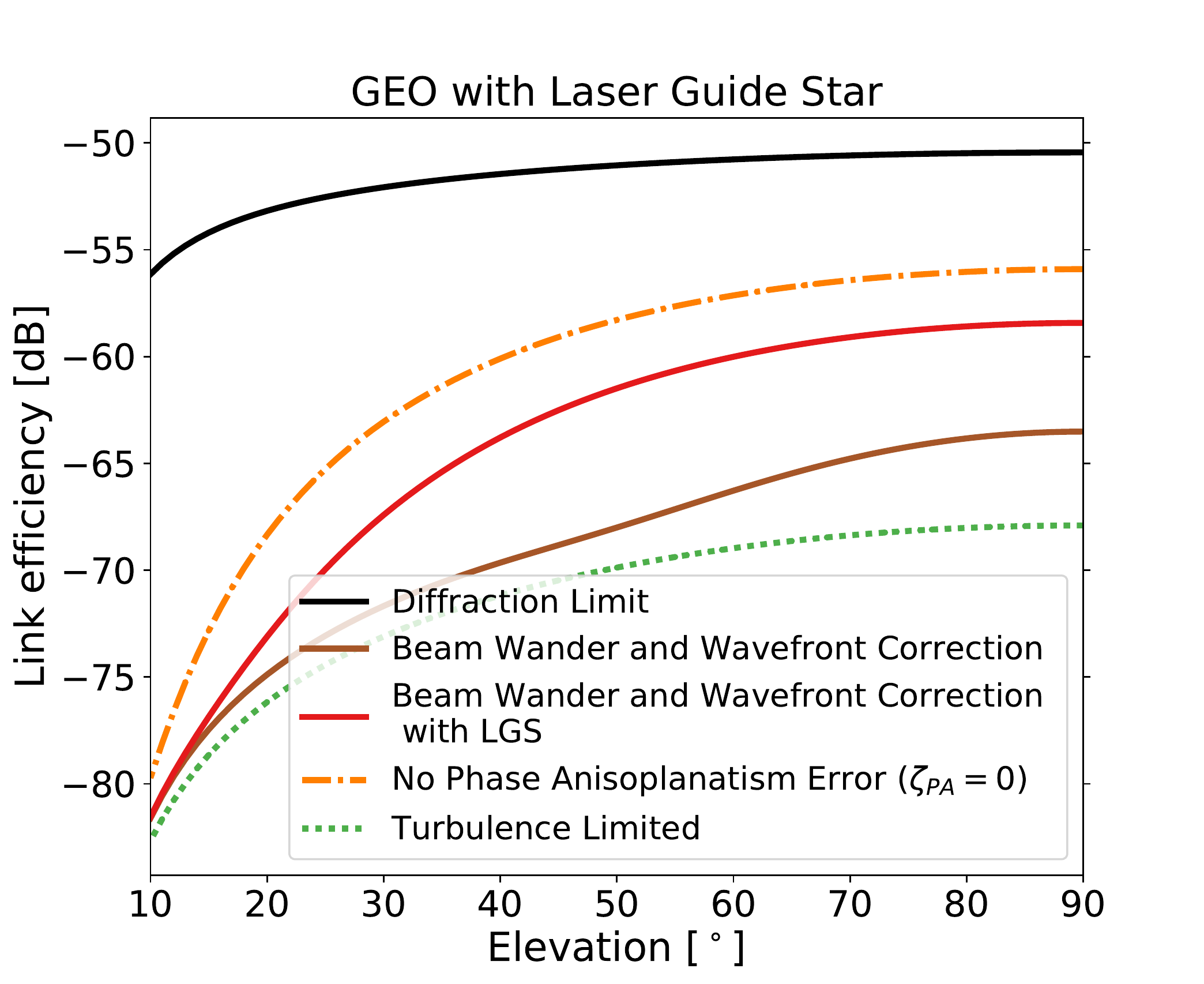}
	\caption{Predicted link efficiency as a function of elevation angle for a geostationary orbit and using a laser guide star. Curves follow definitions in Figure~\ref{fig:LGS}.
	In this HV 5-7 atmosphere model, AO even without LGS becomes more significantly impactful at higher elevations (above \SI{50}{\degree}), in contrast to the LEO case.
	With LGS, an overall improvement of \SI{9.5}{\dB} is seen at zenith, compared to the turbulence-limited baseline.}
	\label{fig:GEO}
\end{figure}

We model the case of an uplink to a GEO satellite, with the results shown in Figure~\ref{fig:GEO}. Due to the larger distances, the overall loss is much higher than for a LEO satellite. It can be seen that the anisoplanatism phase error is still the dominating term---this is because point-ahead, still necessary due to the rotating frame, is large enough that the downlink beacon passes through a different portion of atmosphere than the uplink signal. Like LEO, this error is problematic for AO correction (without LGS) at lower elevations, however the effect becomes more muted at elevations beyond \SI{50}{\degree}.

The improvement from incorporating a LGS (also shown in Fig.~\ref{fig:GEO}) is a little less than the LEO case---about \SI{5.1}{\dB} at zenith compared to without LGS---but the improvement at zenith compared to baseline without AO is more pronounced at \SI{9.5}{\dB}. Interestingly, at \SI{45}{\degree} elevation, the improvement compared to without LGS is greater---about \SI{6.3}{\dB}---while the overall improvement compared to without AO is similar at \SI{9.3}{\dB}. Of course, this analysis does not touch on the additional technical challenges facing operation in a geostationary orbit, which include greater radiation exposure, the need for increased light shielding, and higher launch costs. A GEO satellite would, however, require only static pointing at the transmitter, and thus error from pointing and tracking would be less than for a LEO satellite.

\section{Conclusion}
\label{sec:conclusion}

We have constructed a theoretical model to simulate the effects of atmospheric turbulence on the performance of an uplink from an Earth ground station to a satellite, for purposes where the time-averaged total received optical power is the key parameter, such as QKD. The model incorporates the effects of tracking bandwidth, anisoplanatism, atmospheric turbulence strength, transmitter and receiver size and efficiencies, and technological limitations. As our results show, for the case of a LEO satellite, the most important impact on the link efficiency is the site selection---a \SI{9}{\dB} variation was found between a typical site at sea-level to an excellent astronomical site. In the case where the transmitter location is limited to a particular site, or where further improvement is desired, an AO system can be used to increase the efficiency up to \SI{6.9}{\dB}, under our assumptions of other system parameters. To achieve this, however, it is necessary to correct for anisoplanatism for the AO system to be useful---this can be accomplished by employing a laser guide star. 

Similar anisoplanatism was found when modelling a geostationary satellite---the point ahead necessary for the geostationary link remains large enough that the atmosphere sampled by the downward facing beacon does not correspond well with the signal transmitter upward for elevations below \SI{{\approx}45}{\degree}. For higher elevation angles, anisoplanatism error is still the dominating term but its impact to AO correction is reduced. Use of a laser guide star helps significantly in both cases.

The throughput of QKD protocols depends predominately on the total number of photons measured. The \SI{9}{\dB} improvement from site-selection (if available), as well as the \SI{6.9}{\dB} from AO coupled with laser guide star, would thus deliver a significant increase in the secret key generation rate of the protocol---especially given the characteristics of secret-key-rate formulae for QKD, which imply a favorably nonlinear improvement due to the super-exponential cliff at high losses~\cite{MYMBHJ11}.

\section{Funding}
Canadian Space Agency, Canadian Institute for Advanced Research, Industry Canada, and the Natural Sciences and Engineering Research Council (NSERC).

\section{Acknowledgements}
C.J.P.\ thanks NSERC and the province of Ontario for funding. B.L.H.\ acknowledges support from NSERC Banting Postdoctoral Fellowships.

\bibliographystyle{bibstyle}
\bibliography{aobib}

\end{document}